\documentclass[a4paper]{article}

\usepackage[numbers,sort&compress]{natbib}

\newcommand{\manuscript}{Precipitate size evolution  in an ultrafine-grained magnesium-manganese alloy}

\newcommand{\shortmanuscript}{Mn size evolution}
\newcommand{\shortauthors}{}
\usepackage[numbers,sort&compress]{natbib}

\usepackage[left=2.5cm,right=2.5cm,top=2cm,bottom=2cm]{geometry}
\usepackage{fancyhdr}
	\fancypagestyle{plain}{ %
	\fancyhf{} 
	\lhead{\textit{\shortmanuscript}}
	\rhead{\shortauthors}
	\rfoot{\thepage}
	}
\usepackage{amsbsy} 
\usepackage{graphicx}
	\graphicspath{{./}{./Figures/}} 
\usepackage{subcaption}
\usepackage[version=3]{mhchem} 
\usepackage{multirow} 
\usepackage{multicol}
\usepackage{booktabs}	
\usepackage{setspace}
\usepackage{xspace}
\usepackage{longtable}
\usepackage{comment}
\usepackage{amssymb}
\usepackage{rotating}

\usepackage[plainpages=false,
pdftitle={Precipitate size evolution  in an ultrafine-grained magnesium-manganese alloy},
pdfauthor={J. M. Rosalie and B. R. Pauw and A. Hohenwarter},
pdfsubject={Materials Science},
pdfkeywords={Severe plastic deformation, High-pressure torsion, Ultrafine-grained alloys, Magnesium, Grain-boundary pinning, precipitate growth, interface control, diffusional control},
colorlinks,
debug,
linkcolor=red,
citecolor=red, 
filecolor=red,
breaklinks=false]{hyperref}
\usepackage{doi} 

\title{\manuscript}
\usepackage{glossaries}
\glsdisablehyper 
\setacronymstyle{long-short}

\newacronym{bam}{BAM}{Bundesanstalt f\"{u}r Materialforschung und -pr\"{u}fung, Berlin, Germany.}
\newacronym{bf}{BF}{bright-field}
\newacronym{df}{DF}{dark-field}
\newacronym{ecap}{ECAP}{equal channel angular pressing}
\newacronym{eds}{EDS}{energy dispersive X-ray spectropscopy}
\newacronym{fib}{FIB}{focused ion beam}
\newacronym{grf}{GRF}{growth restriction factor}
\newacronym{haadf}{HAADF}{high angle annular dark-field}
\newacronym{hpt}{HPT}{high-pressure torsion}
\newacronym{hrtem}{HRTEM}{high resolution transmission electron microscopy}
\newacronym{hv}{HV}{Vickers hardness}
\newacronym{lsw}{LSW}{Lifshitz-Slyozof-Wagner}
\newacronym{pals}{PALS}{Positron annihilation lifetime spectroscopy}
\newacronym{pips}{PIPs}{precision ion polishing system}
\newacronym{rlm}{RLM}{reflected light microscopy}
\newacronym{rms}{RMS}{root mean square}
\newacronym{sadp}{SADP}{selected area diffraction pattern}
\newacronym{saxs}{SAXS}{small angle X-ray scattering}
\newacronym{sem}{SEM}{scanning electron microscopy}
\newacronym{spd}{SPD}{severe plastic deformation}
\newacronym{stem}{STEM}{scanning transmission electron microscopy}
\newacronym{tem}{TEM}{transmission electron microscopy}
\newacronym{ufg}{UFG}{ultrafine-grained}
\newacronym{uts}{UTS}{ultimate tensile strength}
\newacronym{xrd}{XRD}{X-ray diffraction}
\newacronym[longplural={regions of interest}, shortplural={ROIs}]{roi}{ROI}{Region of interest}
\newacronym{mouse}{MOUSE}{Methodology Optimization for Ultrafine Structure Exploration}
\newacronym{saxswaxs}{SAXS/WAXS}{Small-/Wide-angle X-ray scattering}
\newacronym{dawn}{DAWN}{Data Analysis Workbench}

\usepackage{authblk}

\date{}
\vspace{6pt}

\author[1]{J. M. Rosalie\thanks{julian.rosalie@bam.de}}
\author[1]{B. R. Pauw}
\author[2]{A. Hohenwarter}
\affil[1]{Bundesanstalt f\"ur Materialforschung und -pr\"ufung,
	Unter den Eichen~87,
	12205,
	Berlin,
	Germany}
\affil[2]{Department of Materials Science, Technical University of Leoben,	Jahnstra{\ss}e 12,	Leoben, 8700, Austria}

\begin{document}
	\maketitle
	\noindent

	\begin{abstract} 
		Precipitate size evolution during room temperature \gls{hpt} of a Mg-1.35wt.\%Mn alloy  was studied using \gls{stem} and \gls{saxswaxs}. The volume fraction of the nm-scale $\alpha$-Mn particles increased with applied strain, however \gls{saxs} indicated that the majority of manganese remained in solution even after 10 \gls{hpt} rotations, indicating that the reaction progress is still limited by the diffusivity of Mn. Analysis of the precipitate size distribution determined that the mean particle size did not increase over the course of \gls{hpt}. This, in combination with the precipitate size distribution suggested that precipitate growth was subject to interfacial rather than diffusional control. 
	\end{abstract}
	
	
	\section{Introduction}
	
	\glsreset{hpt}
	
	\Gls{hpt} is one of several \gls{spd} techniques which are employed to study the microstructure and mechanical properties of metals and alloys. In pure metals, such deformation results in extensive grain refinement, down to the \gls{ufg} or even nanocrystalline regime, with accompanying increases in strength etc. 
	However, few engineering applications rely on the use of pure elements, and in multi-component systems the high defect density\cite{SauvageWetscher2005} and accelerated diffusion\cite{SauvageWetscher2005} can also have a profound influence on the formation of secondary-phases. 
	
	The interplay between \gls{spd} and precipitation is complex. The accelerated diffusion operating during \gls{spd} can result in the formation of additional phases at much faster rates and/or lower temperatures than under conventional conditions\cite{SauvageEnikeev2014}.
	Equilibrium phases may form directly, bypassing the normal precipitation sequence\cite{Duchaussoy2020}. Rapid precipitation of intermetallic phases can occur during deformation, even in the absence of external heating. Such ``dynamic'' precipitation has been reported during \gls{ecap} and/or \gls{hpt} deformation of Al-Cu \cite{Nasedkina2017,Hohenwarter2014}, Al-Ag \cite{Ohashi2006}, Al-Zn \cite{Chinh2020} and Al-Zn-Mg-Cu \cite{HuMa2013,Duchaussoy2020} alloys.
	
	Conversely, forced intermixing can increase the effective solubility limit and, like mechanical alloying via ball milling, can result in single-phase alloys in otherwise insoluble systems, precluding precipitation entirely 	\cite{SauvageWetscher2005,KormoutPippan2016}. In addition, when precipitates are present, they, like other 	second phases, are also vulnerable to deformation during \gls{spd}. This can affect both the precipitate size and morphology. Dynamically-precipitated $\theta$ phase particles in Al-Cu alloys underwent a reduction in size (from an initial value of 70 to around 40\,nm) on further deformation\cite{Nasedkina2017,Hohenwarter2014} due to fragmentation and/or plastic deformation of the particles at higher strains\cite{Nasedkina2017,Hohenwarter2014}. 
		
	A recent investigation by the authors into grain refinement of a Mg-Mn alloy  found that nanometer-scale Mn particles formed dynamically during \gls{hpt} \cite{Rosalie2025}. This investigation was motivated by the fact that dynamically-precipitated Mn particles are known to act as pinning particles during high-temperature creep of Mg-Mn alloys.
	Furthermore, Mn is one of the few common alloying elements used in Mg which is suitable for biomaterials applications as a  biodegradable implant material\cite{GuZheng2010,ValievEstrin2016,YangYoon2015,GutierrezPua2023}.
	
	The 2--5\,nm diameter Mn particles are thought to have stabilised the grains against coarsening\cite{Rosalie2025}, allowing the alloy to reach grain sizes of 140\,nm, much finer than pure Mg \cite{Edalati2011,Sulkowski2020,QiaoZhao2014} and comparable with several, more heavily-alloyed commercial magnesium compositions\cite{XiaWang2005,Harai2008,XuShirooyeh2013,Silva2019,Yu2015}.	Surprisingly, \gls{tem} images of the Mn particles showed no obvious change in dimension, even after extended deformation, while the matrix grains underwent coarsening, growing to 240\,nm after 10 rotations of \gls{hpt}. The present work sets out to clarify the process of precipitate growth and thus clarify the reasons for continued grain growth during deformation.

This work describes a detailed examination of the evolution of the particle size distribution in the Mg-Mn alloy during extended \gls{hpt} deformation. It combines the use of atomic-contrast \gls{stem} and \gls{eds} to determine the spatial distribution, morphology and chemistry of the particles. This information facilitated the development of a physically representative model for the particle size and morphology \cite{RosaliePauw2014} for use in SAXS. This enabled \gls{saxs} to accurately measure particle the size distribution as a function of the imposed strain. The experimental data is compared with predictions from models for precipitate growth mechanisms, and the role of the particles in stabilising the magnesium grains is discussed.

	\section{Experimental}

	All experiments were carried out using a commercial grade M1A alloy containing of nominal composition Mg-1.35wt.\%Mn,  sourced from Magnesium Elektron. The full compositional analysis is available \cite{Rosalie2025}, and shows that no additional elements were present at levels greater than 0.01wt.\%.

\begin{figure}[hbtp]
	\centering
	\includegraphics[width=0.48\textwidth]{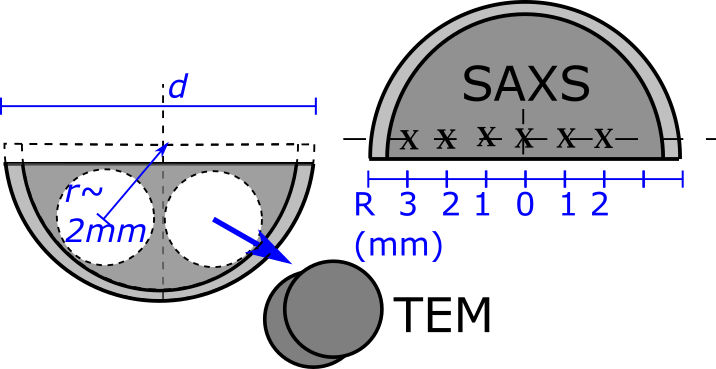}
	\caption{A schematic of the \gls{hpt} disc geometry and sample preparation. SAXS measurements were made at radii between 0 and 3\,mm. The centre of the \gls{tem} foils corresponds to a radial distance of approximately 2\,mm.  
		 \label{fig:sample-geometry}}
\end{figure}

	Preparation of the \gls{hpt} discs is described in more detail elsewhere \cite{Rosalie2025}. Discs were sectioned, as illustrated in Fig~\ref{fig:sample-geometry} to provide one sample for position-resolved \gls{saxs} and two \gls{tem} foils per \gls{hpt} disc. Samples for \gls{saxs} were ground and polished to a thickness of 50\,$\mu$m. The preparation of the \gls{tem} foils was performed using mechanical grinding and ion polishing and has been described elsewhere \cite{Rosalie2025}.  

	\Gls{stem} observations and \gls{eds} maps were obtained using a ThermoFisher Scientific Talos F200S microscope operating at 200\,kV. Particles were characterised by \gls{haadf} \gls{stem}, using an in-house ImageJ (version 1.53)\cite{Schneider2012} script  to automatically segment the images and measure the size and shape parameters. A minimum of 2000 particles were included in the analysis for each deformation condition. 
	
	\Gls{eds} maps were analysed with scripts developed using the HyperSpy (version: 2.1.1)\cite{hyperspy_v2_1_1} and Exspy (Version 0.2.1)\cite{exspy_v0_2_1} Python libraries. This involved comparing the X-ray intensity for one or more atomic species in different \glspl{roi}, for example intragrannular regions and nearby grain boundaries. The X-ray intensity for each species within a pair of \glspl{roi} was subjected to a two-sided Student's t-test of the hypothesis that the regions did not differ in composition. This was considered falsified if the probability of this, was less than 0.01 i.e. when there was a 99\% probability that significant differences existed. This method enabled comparisons to be made regions of differing sizes with low X-ray counts. 
	
	 \Gls{saxswaxs} measurements were obtained using the \gls{mouse} \cite{Smales_2021}, at radial distances to 3\,mm, as illustrated in Fig.~\ref{fig:sample-geometry}.  X-rays were generated from a microfocus X-ray tube, followed by multilayer optics to parallelize and monochromatise the X-ray beams to wavelengths of Cu K$\alpha$ ($\lambda$ = 0.154\,nm).
	 Scattered radiation was detected on an in-vacuum Eiger 1M detector (Dectris, Switzerland), which was placed at multiple distances between 55--2507 mm from the sample. The incident beam diameter was estimated at 0.49$\pm$0.05\,mm from multiple direct-beam images. The resulting data has been processed and scaled to absolute intensity using the \gls{dawn} software package\cite{Filik2017} in a standardized complete 2D correction pipeline with uncertainty propagation\cite{Pauw2017}.

	\section{Results}
	
	\begin{figure*}
		\hfill\ 
		\hfill\
		\subfloat[0 rotations \label{fig:mgmn-stem-comp}]{\includegraphics[width=0.32\linewidth]{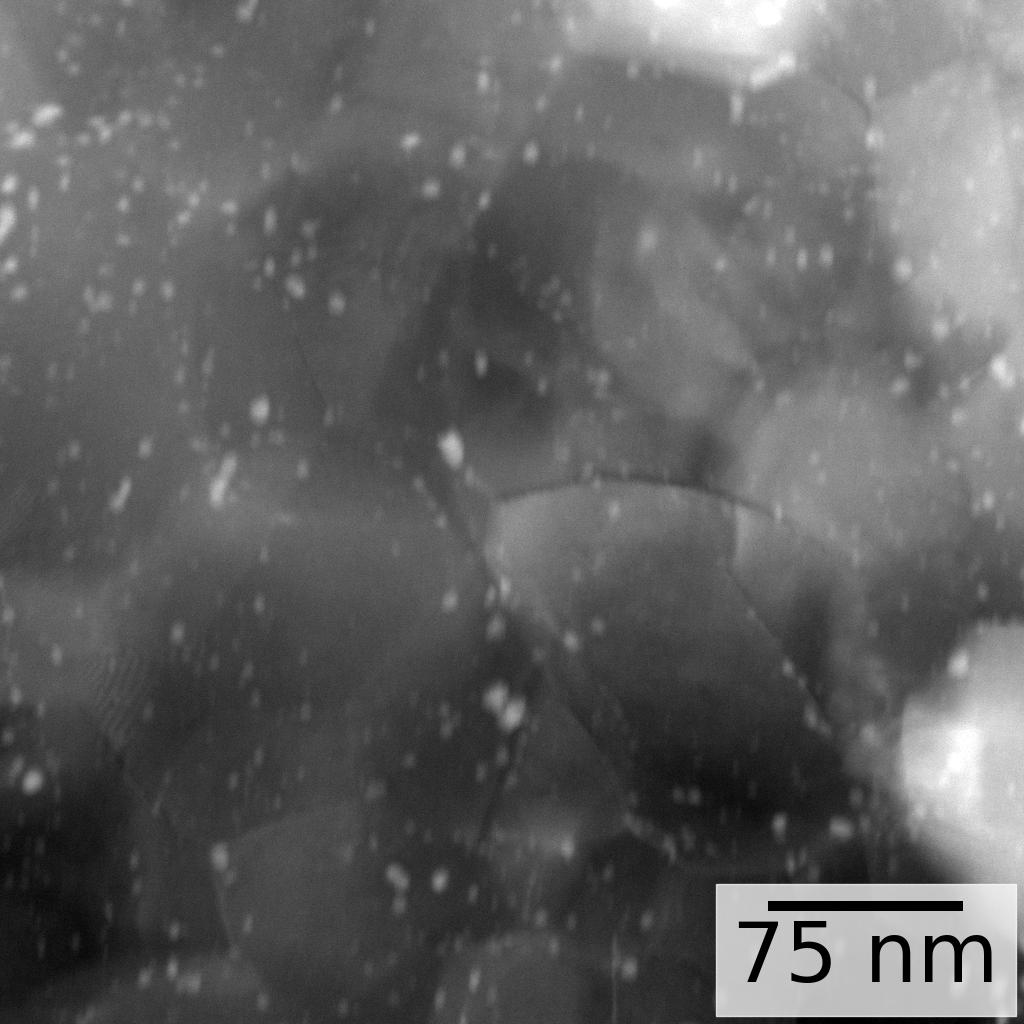}}
		\hfill\
		\subfloat[0.5 rotations \label{fig:mgmn-stem-0p5}]{\includegraphics[width=0.32\linewidth]{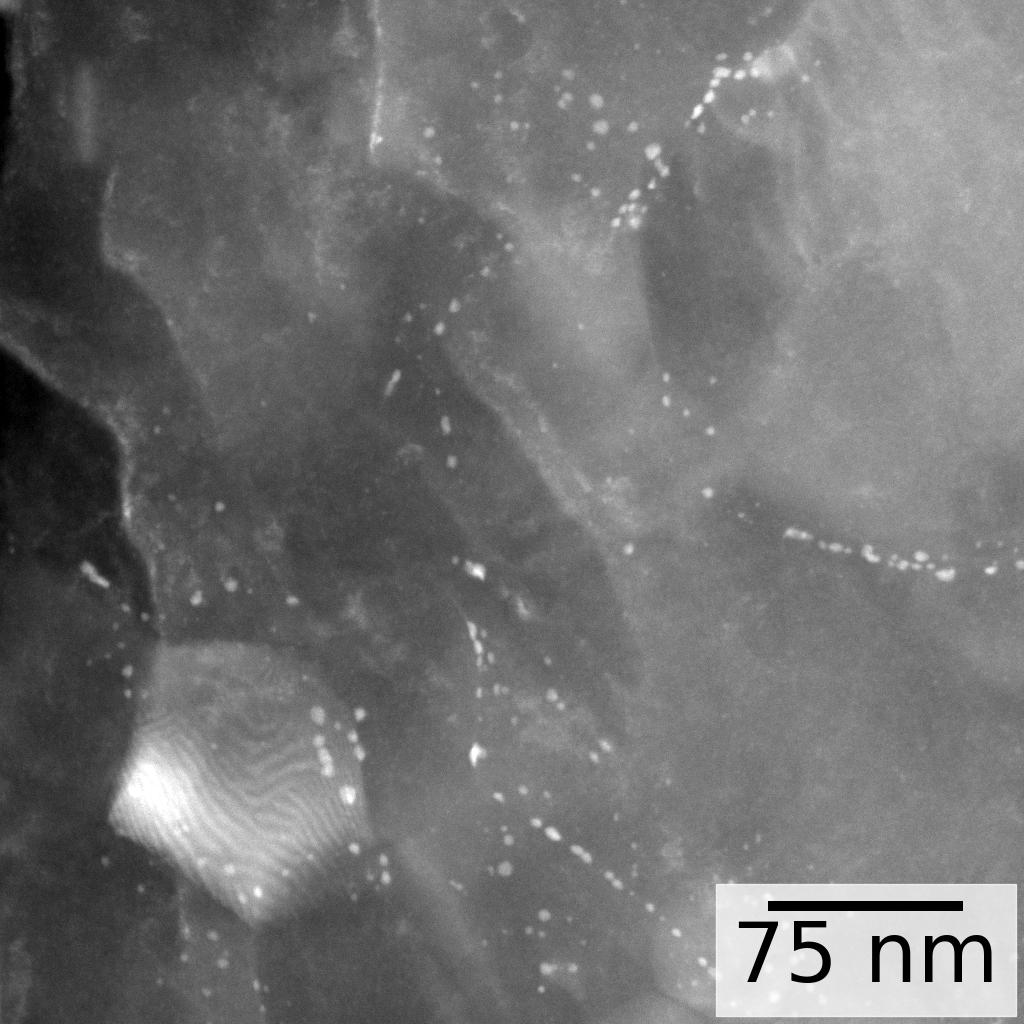}}
		\hfill\ 
		\subfloat[5 rotations \label{fig:mgmn-stem-5}]{\includegraphics[width=0.32\linewidth]{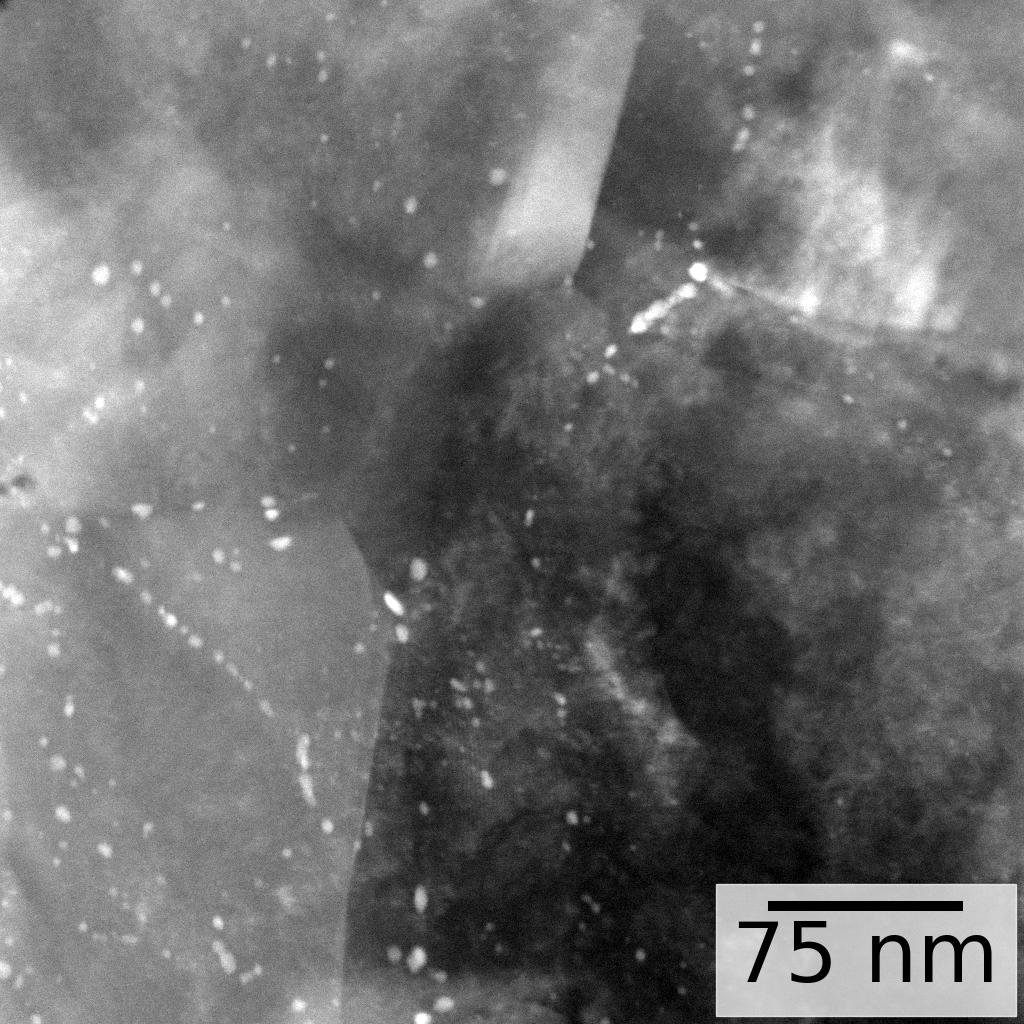}}
		\hfill\  
		\caption{\Gls{haadf}-\gls{stem} images post-\gls{hpt} Mg-Mn, for (a) 0 rotations (i.e. compression only), (b) 0.5 rotations and (c) 5 rotations. Nano-scale Mn particles are concentrated along the grain boundaries.\label{fig:mgmn-stem-summary}}
	\end{figure*}

	\subsection{Post-deformation microstructure}
	
\begin{figure}[hbtp]
	\begin{center}
		\hfill
		\subfloat[\label{fig:tem-particle-size-mean}]{\includegraphics[width=0.4 \textwidth]{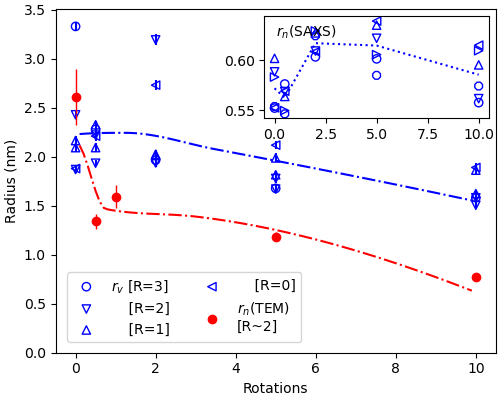}}		
		\hfill
		\subfloat[\label{fig:tem-particle-aspect-ratio}]{\includegraphics[width=0.48\textwidth]{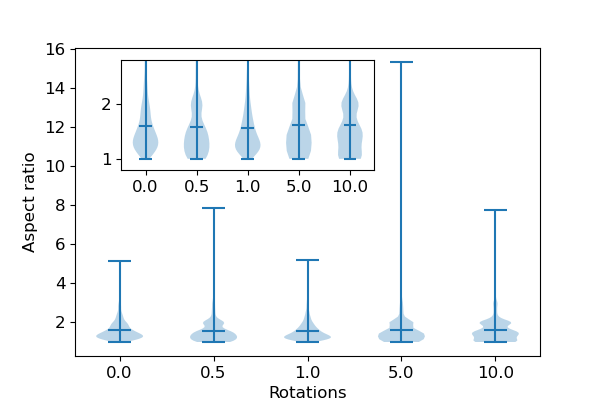}}
		\hfill\ 
			\caption{Development of the Mn particle distribution 		
			(a) Mean particle radius. Volume-weighted values for each \gls{saxs} radial position are indicated by open symbols (blue in the colour version online). Number-weighted values from \gls{haadf}-\gls{stem} are shown by filled symbols (red in the colour version online).
			The inset shows the \gls{saxs} data, converted to number-weighted radii.			
			Curves are provided as a guide for the eye, only. 
			(b) Violin plot showing the aspect ratio of the particles as measured by \gls{haadf}-\gls{stem}. The horizontal lines indicate the maximum, mean and minimum values, respectively.
			\label{fig:particle-summary-tem}}
		\end{center}
\end{figure}

\Gls{haadf}-\gls{stem} images of the post-\gls{hpt} material showed ultrafine magnesium grains, with bright (i.e. high atomic-number) particles concentrated along the grain boundaries. Figure~\ref{fig:mgmn-stem-summary}
shows representative images for 0.5, 5 and 10 rotations, respectively. Precipitate size measurements revealed an unexpected reduction in  the particle radius over the course of the deformation process, from 2.6\,nm after 0 rotations (ie. in \gls{ufg} regions formed after compression), to 0.8\,nm after 10 rotations (Indicated by filled circles and coloured red in the online version) in   Fig.~\ref{fig:tem-particle-size-mean}.) The most substantial change occurred between 0.25 and 1 rotations  after which the rate of reduction appeared to stabilise. 

The bright particles were relatively isotropic, as can be seen in Fig.~\ref{fig:tem-particle-aspect-ratio}, which presents the aspect ratio distributions as a violin plot. The horizontal lines indicate the maximum, mean and minimum values, respectively, as determined from the \gls{haadf}-\gls{stem} images. Despite the presence of some outliers, the aspect ratios were strongly centred at ~1.25, with a weaker cluster at $\sim$2 particularly after 5 and 10 rotations. The mean aspect ratio did not change significantly during deformation, varying unsystematically between 1.58 (0.5 rotations) to 1.62 (5 rotations). 
	
	\subsection{Compositional mapping}
\begin{figure}
	\centering
	\hfill
	\subfloat[\label{fig:0016---20240925-mgmn-1rotn-d2-si-360-kxstem-image3-a}]{\includegraphics[width=0.48\textwidth]{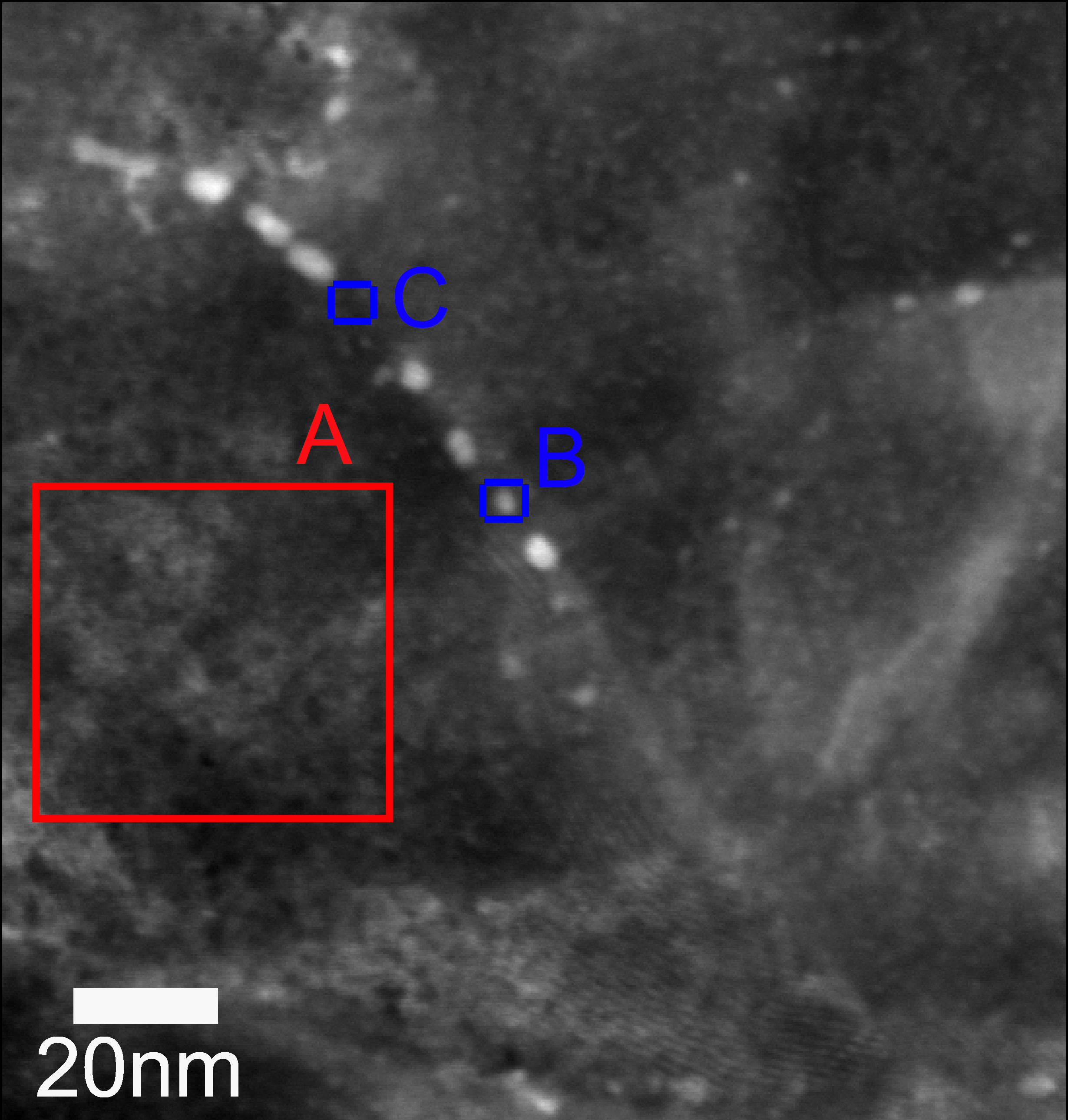}}
	\hfill
	\subfloat[\label{fig:0016---20240925-mgmn-1rotn-d2-si-360-kxstem-image3-c}]{\includegraphics[width=0.48\textwidth]{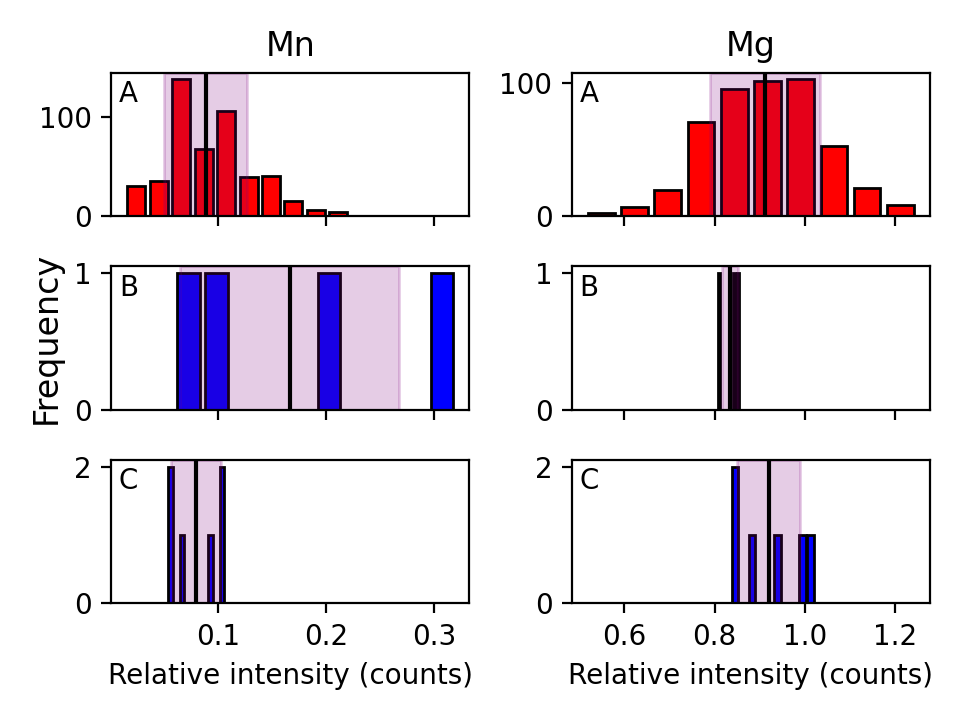}}
	\hfill\ 
	\caption{(a)\gls{haadf}-\gls{stem} image obtained in a sample after 1.0 rotations of \gls{hpt}. 
		Regions of interest (A,red in the online version) in the grain, (B, blue in the online version) around a grain boundary particle and (C, blue) along the grain boundary are indicated. Histograms of the relative \gls{eds} intensity for Mn and Mg for each region are shown in (b).
		The shaded band (purple in the colour version online) indicates a width of 1 standard deviation above and below the mean. The relative intensity in region $B$ is substantially higher than in the grain ($A$) or at the boundary itself ($C$).}
	\label{fig:0016---20240925-mgmn-1rotn-d2-si-360-kxstem-image3}
\end{figure}

\Gls{eds} mapping confirmed that Mn is strongly localised in the grain-boundary precipitates, but yielded no evidence of solute segregation at the grain boundaries themselves. 
Figure~\ref{fig:0016---20240925-mgmn-1rotn-d2-si-360-kxstem-image3-a} shows precipitates along a grain boundary in a sample deformed by one rotation of \gls{hpt}. \Glspl{roi} (A) in the grain interior, (B) centred on a precipitate and (C) on the grain boundary between precipitates are indicated. The histograms for the X-ray intensity (mean counts per pixel) in regions A and B are provided in Fig.~\ref{fig:0016---20240925-mgmn-1rotn-d2-si-360-kxstem-image3-c}, and show a dramatic enhancement of the X-ray signal for Mn, together with a decrease for Mg. A Student's $t$-test on the Mn K$\alpha$ intensity distribution in regions $A$ and $B$ 
indicates that they differ in composition with $>99\%$ probability. In contrast, region $C$ shows a slight (but not statistically significant) decrease in Mn.
Numerous such analyses for samples deformed by either 1.0 or 10 rotations of \gls{hpt} determined that statistically significant increases in Mn content could be reliably detected for particles of diameter $\ge$2\,nm, whereas no such enhancement was detected in similarly-sized \glspl{roi} along grain boundaries.

\subsection{SAXS particle size distributions}

\begin{figure*}
		\centering
		\includegraphics[width=0.9\textwidth]{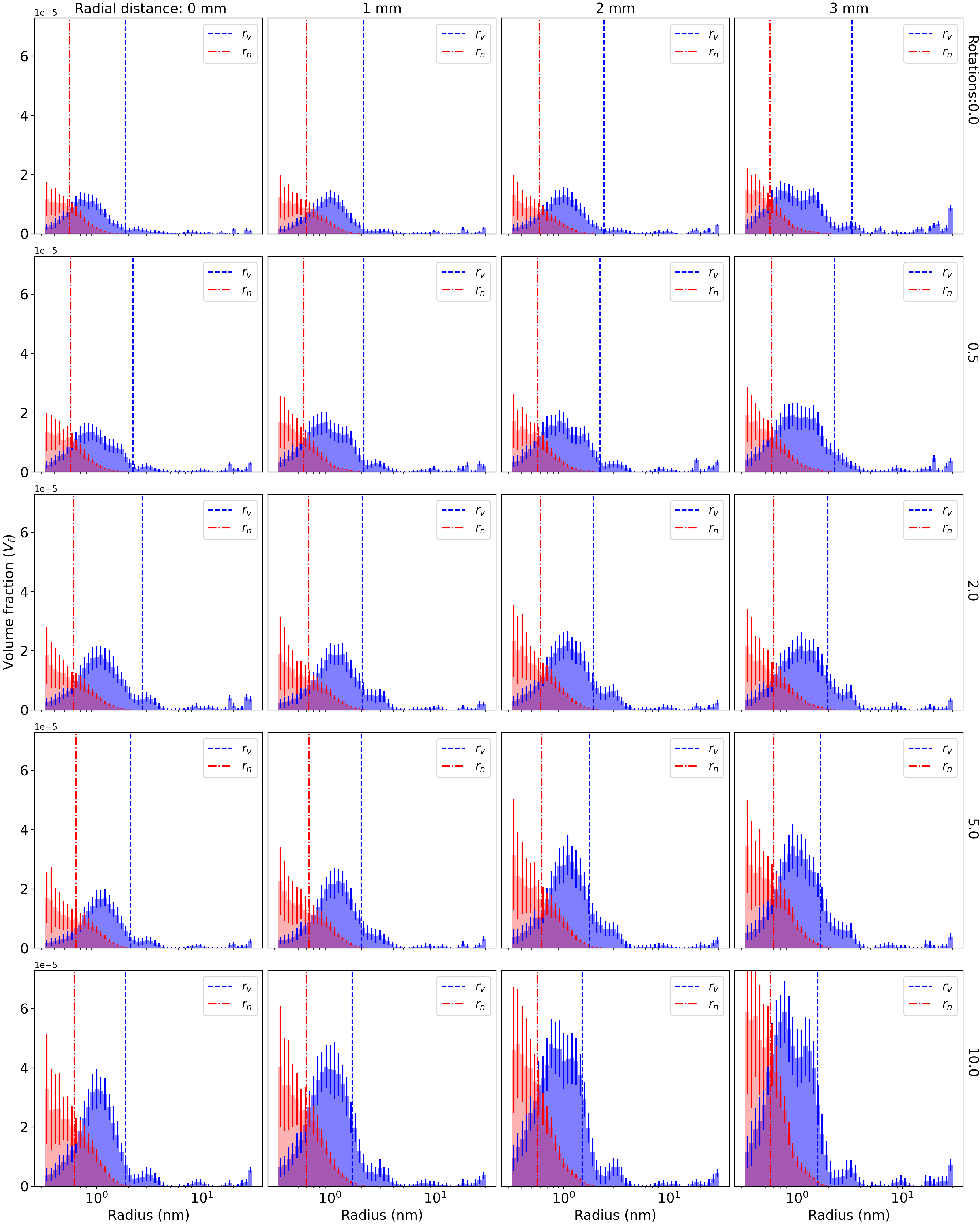}
		\caption{Particle size distributions from \gls{saxs} measurements. Darker histograms (blue in the colour version online) indicate the volume-weighted particle size distributions, while the lighter (red in the colour version) histograms indicate the number-weighted distributions. The standard deviation for each histogram is shown by solid lines, and dashed vertical line shows the mean particle radius. The volume fraction of Mn particles increases with applied \gls{hpt}  strain, both (i) from left to right with increasing radial distance, from the disc centre (left in the figure) to the rim (right) and (ii) from top to bottom with increasing number of \gls{hpt} rotations.
			($\overline{r}_v$). 	\label{fig:saxsradiushistogramsb}}
	\end{figure*}

\Gls{saxs} curves showed an increase in scattering cross-section for scattering vectors ($Q$) in the 0.3--30\,nm range. Using the \gls{haadf}-\gls{stem} data as a basis, the particle size distribution could be determined with McSAS3 \cite{Pauw2013a,Bressler2015}, modelling the precipitates as spherical $\alpha$-Mn particles\cite{Rosalie2025}. The fitted particle size distributions are shown in Figure~\ref{fig:saxsradiushistogramsb} and show a broad peak, centred on a  radius of 1--2\,nm, with a low fraction of particles with radii up to 30\,nm. 

The vertical axis in the histograms indicates the net volume fraction within each size bin and shows a gradual increase with increasing applied strain, both (i) from left to right with increasing radial distance (from the centre at $R$=0\,mm, left) to the rim (right) and (ii) from top to bottom with the number of rotations. The most significant change in volume fraction occurs for particles in the 1--2\,nm size range.

The mean particle radii  are plotted in 	Fig~\ref{fig:tem-particle-size-mean} (shown in blue in the colour version online) indicating the number of rotations, and the distance from the centre of the \gls{hpt} disc. With the exception of the outermost part of the disc (radial distance of 3\,mm), all positions showed an maximum in particle radius at 0.5 rotations, and, like the values from \gls{stem} imaging, a gradual decline thereafter. It is thought that at this outermost position position the X-ray beam might impinge on the rim of the disc, where the thickness changes rapidly and the strain condition is quite complex. 

The	inset in Fig~\ref{fig:tem-particle-size-mean} shows the \gls{saxs} values converted to number-weighted radii. These values are considerably lower than the values obtained from \gls{stem} and rather than decreasing, show a very minor upward trend, increasing from 0.418\,nm after compression to 0.423\,nm after 10 rotations. The discrepancies between the different measures of particle size  yield valuable information about the precipitate growth and are covered in greater detail in the discussion.

\begin{figure}[hbtp]
		\begin{center}
			\hfill\ 
			\subfloat[\label{fig:SaxsVolumeStrain}]{\includegraphics[width=0.4\textwidth]{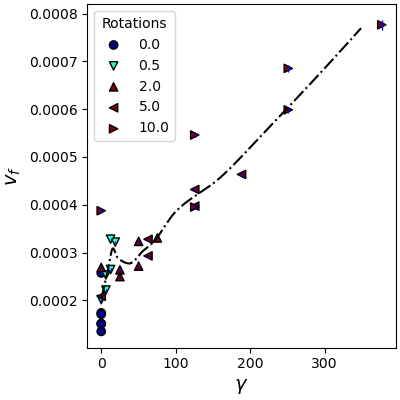}}		
			\hfill
			\subfloat[\label{fig:SaxsNumberStrain}]{\includegraphics[width=0.4\textwidth]{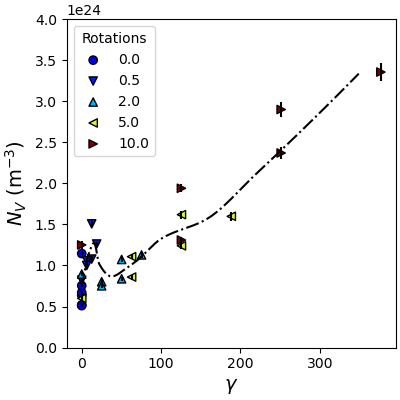}}
			\hfill\ 
			\caption{Development of the Mn particle distribution 					
				Curves are provided as a guide for the eye, only. 
				(c) Total particle volume fraction (from \gls{saxs}) versus equivalent strain.
				(d) Particle number density versus equivalent strain.
				\label{fig:particle-summary-saxs}}
		\end{center}
	\end{figure}

	The volume fraction (Fig.~\ref{fig:SaxsVolumeStrain}) and number density (Fig.~\ref{fig:SaxsNumberStrain}) of particles increased with the applied strain. The total volume fraction, increased rapidly to 0.025\% followed by a gradual consistent increase with applied strain for $>$ 2 rotations. The maximum value 0.077v/v\% was still well below the stoichiometric amount (0.31\%) expected for complete precipitation of Mn in the alloy. The number density was determined using the \gls{saxs} (number-weighted) radii and shows a very similar trend to the volume fraction with a rapid jump in number density, followed by roughly linear increase with the strain for numbers of rotations of 2 or greater. 
	
\section{Discussion}
	
Before addressing questions about precipitate growth in the \gls{hpt}-deformed Mg-Mn alloy, it is important to consider the differences between the techniques employed to measure particle populations. Imagining methods like \gls{tem}/\gls{stem} direct detect a particle based on contrast differences with the surrounding material, providing a number-weighted particle size based on individual measurements. This has the advantage of proving shape descriptors, such as the aspect-ratio, but is inherently limited to small sample sizes. The choice of imaging technique is important, and \gls{haadf}-\gls{stem} imaging was used here due to the large difference in atomic number between Mg and Mn, and hence strong contrast between phases. However, the ease of distinguishing the particles from the matrix also depends on the particle size and relative thickness of the matrix, and particles of the scale present in this alloy can be challenging to reliably detect. 

\gls{saxs} probes a much large volume, and the signal is proportional to the volume of scattering sites within a particular size range.
\Gls{saxs} is able to detect smaller scattering sites, which may not be consistently visible in \gls{tem}/\gls{stem}, even when there is good atomic contrast between matrix  and precipitates. 
The resulting particle size distributions effectively appear in the scattering signal as volume-weighted information, but although a transformation to a number-weighted size distribution is possible, results for the smallest components will be subject to very high uncertainties. 
In this case, \gls{saxs} may suggest the presence of a proportion of sub-nm scatterers, possibly atomic clusters or precipitate nuclei, which are absent from the \gls{stem} images. Therefore, although differences between the values obtained from imaging and scattering methods will be noted, issues relating to the precipitate growth, solute segregation etc. will rely on the numerical values provided by \gls{saxs}.

The reduction in \gls{saxs} volume-weighted particle radii with increasing deformation  (Fig~\ref{fig:tem-particle-size-mean}) actually indicates an \textit{increase} in the proportion of smaller particles, as can be seen in the lower-right section of Fig~\ref{fig:saxsradiushistogramsb}. A similar trend is evident in number-weighted values from \gls{stem}, although the values are lower, since the volume-weighting highlights the contribution of the larger particles. Converting the \gls{saxs} data to number weighted values, is particularly useful however. Firstly, the radii are much lower, as it neither emphasises the large particles, nor fails to detect the smaller ones. Secondly the scatter is greatly reduced compared to the volume-weighted \gls{saxs} values, suggesting that this is due to the presence of a small population of larger, outlier particles. Finally, it shows that there is, at most, a very minor increase in the particle size during deformation. This is a somewhat surprising result when compared with other studies on \gls{spd} of precipitate-containing systems, particularly those in which dynamic precipitation occurs and this point in taken up later in this discussion. 

\subsection{Solute redistribution during deformation}

\Gls{hpt} deformation process results in a redistribution of solute due to the continual generation of vacancies during \gls{spd} deformation and their migration \cite{Duchaussoy2020,SauvageEnikeev2014}. This will lead to a net solute flux towards the grain boundaries. Depending on the alloy system, this can result in solute segregation and wetting of the grain boundaries, and/or precipitation at the boundaries and triple points. 

Theoretical studies predict an energetic advantage for grain boundary segregation for Mn in Mg \cite{Somekawa2018}, and several reports on binary \cite{SomekawaBasha2018} and ternary Mg-Mn-Zr \cite{Somekawa2019}, Mg-Mn-Bi\cite{Somekawa2019} and  Mg-Mn-Nd \cite{deOliveira2021,WangCepeda2020} alloys support this. However, the \gls{haadf}-\gls{stem} images and \gls{eds} mapping  (Fig.~\ref{fig:0016---20240925-mgmn-1rotn-d2-si-360-kxstem-image3}) in the present work show no evidence of solute enrichment. This is is despite the low volume fraction of Mn particles which have precipitated (Fig.~\ref{fig:SaxsVolumeStrain}), and  implies that a large amount of Mn has remained in solution. Testing this hypothesis is important in order to understand the growth kinetics of the particles and their role in grain-size stabilisation. 

A rudimentary model can be constructed using similar assumptions to those which Sauvage et. al.\cite{SauvageEnikeev2014} employed to estimate the effective temperature during \gls{hpt} deformation of an Al-Mg alloy. These authors assumed that diffusion follows a random walk model, in which case the diffusion distance, $\lambda_{eff}$, can be found from the equation:

\begin{equation}
\lambda_{eff} = (6Dt)^{1/2} 
\label{eqn-lamba-effective}
\end{equation}
where $D$ is the diffusivity and $t$ the reaction time and assuming that the diffusion distance,  $\lambda \sim \lambda_{eff}$ and that the diffusion distance should be of the order of  1/4 of the grain size\cite{SauvageEnikeev2014}. 
This assumed extensive precipitation and the lower precipitate proportion here suggests that the diffusion distance in Mg-Mn is lower than in Al-Mg. Instead we can use a simple, geometric model, taking into account the precipitate volume fraction obtained from \gls{saxs}.

We begin by assuming a spherical grain of radius, $R$, and that diffusion to the grain boundaries results in a depleted layer of width, $ \lambda = kR$, where $k$ is a constant.  

\begin{eqnarray}
V*  & = &\frac{4}{3}\pi(R-\lambda)^3 \\
& = & \frac{4}{3}\pi  (1-k)^3R^3
\end{eqnarray}

\begin{figure}[hbtp]
	\begin{center}
	\includegraphics[width=0.75\textwidth]{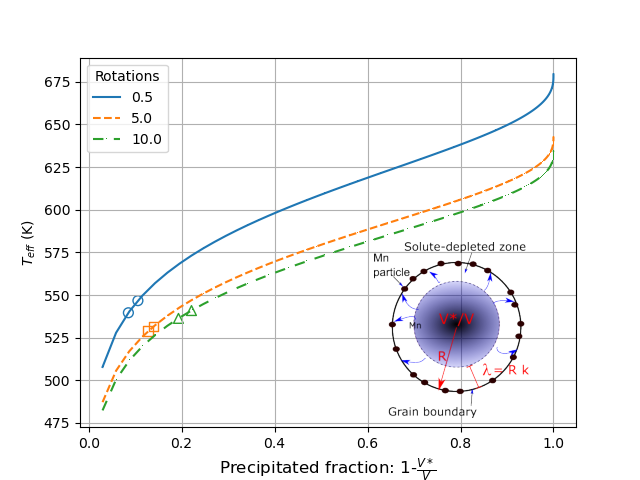}
	\caption{Schematic of the relationship between the extent of precipitation and the effective temeperature duing \gls{hpt} deformation.\label{fig:diffusiondistance}} 
	\end{center}
\end{figure}

The solute level in the remainder of the grain remains constant, and the proportion of solute remaining is then $V*/V$ which is equal to   $(1-k)^3 $. Conversely the relative volume fraction fraction of precipitates is:
\begin{equation}
\frac{V_f}{V_f^{max}} = 1 - (1-k)^3
\end{equation}
This model is illustrated in the inset in Fig.~\ref{fig:diffusiondistance} and shows the solute depleted layer and a number of boundary particles. 

The effective temperature can be estimated using the \gls{saxs} data for a radial distance of 2\,mm, (which corresponds most closely to the \gls{stem} observations), and the grain size values reported previously for this alloy\cite{Rosalie2025} to determine $k$ for a given number of rotations. Diffusion constants of  D$_0$ = 0.76 x 10$^{-4}$s$^{-1}$ and $Q$ = 176 (kJ mol$^{-1}$) were taken taken from the work of Fujikawa \cite{Fujikawa1992}. The reaction time in Eqn.~\ref{eqn-lamba-effective} was determined by the number of rotations and the rotation rate of 1 revolution per minute\cite{Rosalie2025}.

The curves in Fig.~\ref{fig:diffusiondistance} indicate $T_{eff}$ for a a given fraction of the total solute which is partitioned to precipitates. For a given number of rotations, an increase in the effective temperature leads to a rapid increase in the amount of precipitation. In the present case, however, the experimental data shows a low, but increasing precipitate volume fraction with the number of rotations, reaching 0.21\% after 10 rotations. This corresponds to an effective diffusion distance of only $\sim$20\,nm.  Given the assumptions inherent in the model, the $T_{eff}$ is quite consistent, ranging between 530--547\,K. Despite this somewhat high effective temperature, it does suggest that even under these deformation conditions, precipitation will proceed sluggishly.

From an examination of the \gls{stem} images (Fig.~\ref{fig:mgmn-stem-summary} and \ref{fig:0016---20240925-mgmn-1rotn-d2-si-360-kxstem-image3} ), it is clear that some, probably many, particles would fall within this range of their neighbours, even without taking into account short-circuit diffusion along the grain boundaries. The continued presence of particles in such proximity to one another raises questions about diffusion between particles and its effect on particle growth, which are taken up in section~\ref{sec-precipitate-growth}.
	
\subsection{Grain growth stabilisation}

The recent examination of \gls{hpt} of Mg-Mn reported a much finer minimum uniform grain size of 140\,nm\cite{Rosalie2025} than has been achieved through \gls{spd} of pure magnesium, despite the low alloying content. In addition, the alloy did not develop the bimodal grain structure which is known to develop during \gls{hpt} deformation of the pure metal. Both observations indicate that Mn plays a role in restricting dynamic grain growth during deformation.

Grain boundary stabilisation during \gls{spd} is usually classified as either thermodynamic (grain boundary segregation) or kinetic (Zener or particle pinning) in origin. Although there is a tendency for grain boundary segregation of Mn to occur in Mg \cite{SomekawaKinoshita2017, SomekawaSingh2016,Somekawa2018} , it is possible to exclude its effect in the present work, based not only on the \gls{haadf}-\gls{stem} and \gls{eds} data but also on the extent of grain refinement. 

For segregation to provide effective resistance to grain growth, it is considered that sufficient solute must be present to wet all all of the boundaries.  The critical concentration for which, $C$ is given by: 
 \begin{equation}
 C = \frac{3\delta}{2r} \label{eqn-critical-grain-boundary-coverage}
 \end{equation}
 where $r$ is the grain radius, and $\delta$ is the grain boundary width, taken to be 2$\mathbf{b}$.
 ($\mathbf{b}= 3.14\times 10^{-10}$m for Mg).

\begin{figure*}
	\centering
	\includegraphics[width=10cm]{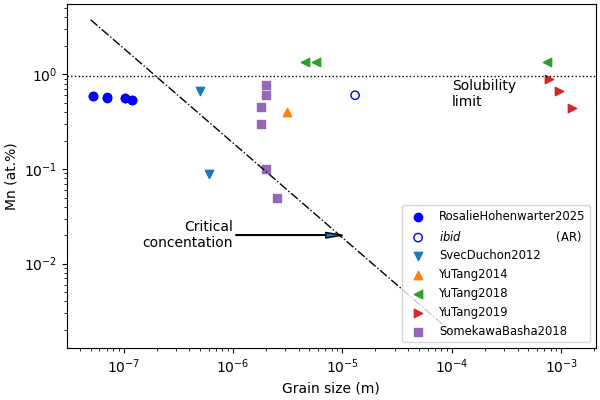}
	
	\caption{Critical solute level for grain boundary coverage. ``AR" and ``HPT" indicate the as-received and \gls{hpt} samples, respectively. All \gls{hpt}-deformed samples have grain sizes below the critical value required for grain size stabilisation. Additional data from the literature on binary Mg-Mn alloys is included for comparison.
		 \label{fig:critcalgrainsegregation}}
\end{figure*}

The critical concentration is plotted for various Mg-Mn alloys  in Fig.~\ref{fig:critcalgrainsegregation}.
The solute content in deformed M1A was calculated for residual solute values from \gls{saxs} combined with the previously-reported grain sizes \cite{Rosalie2025}. In all \gls{hpt}-deformed samples in this work  (filled blue circles) the grain size is reduced to a level where the amount of solute is well below the critical concentration. Indeed for $\le$2--5 rotations the required solute level exceeds the solubility limit. 
Additional data is included from previous studies on Mg-Mn binary alloys\cite{YuTang2014, YuTang2018,SvecDuchon2012,YuTang2019,SomekawaBasha2018}. With the notable exception of singular samples produced by Somekawa et. al. \cite{SomekawaBasha2018} and \u{S}vec et. al. \cite{SvecDuchon2012} all of these samples fall within the regime where the grain size and solute content would permit complete grain boundary coverage. 

\begin{figure}
	\centering
	\includegraphics[width=7cm]{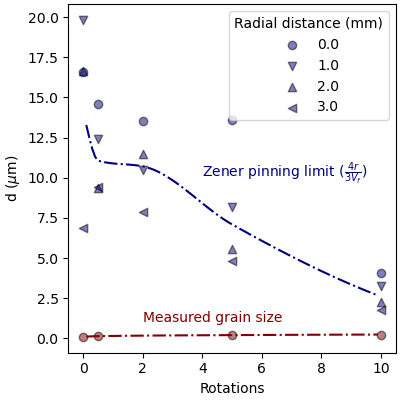}
	\caption{Limiting grain size based on Zener pinning by Mn particles. Even after 10 rotations the grain size is below the limiting value, permitting further grain growth.}
	\label{fig:saxszenerpinning}
\end{figure}

Manganese particles have been shown to act as grain boundary pinning particles\cite{Robson2011,ManoharFerry1998,Koju2016,Qin2023}. 
The effectiveness of Zener pinning is determined by the size and volume fraction of the pinning particles, and the theory predicts a limiting grain size ($D_Z$) where the pinning pressure is sufficient to halt grain growth. For uniformly distributed particles of radius, $r$, present with volume fraction of $V_f$, the limiting grain diameter is given by \cite{Robson2011}:
\begin{equation}
D_Z =  \frac{4r}{3V_f}   
\label{fig:zenerpinning}
\end{equation}
The critical grain diameter was determined from the \gls{saxs} data and is plotted together with the grain size in Fig~\ref{fig:saxszenerpinning}. For consistency, only the \gls{saxs}  data for a radial distance of $\sim$2\,mm are included; which matches most closely with the region examined via \gls{tem}. The grain size is well below the critical grain diameter, which is consistent with the occurrence of grain growth duration during \gls{hpt}, albeit at a sluggish rate. Dynamic precipitation results in decrease in the critical grain diameter with increasing deformation, suggesting that the alloy will at some point reach a dynamic equilibrium and a stable grain size where the precipitate array restricts further grain growth. It should be noted that  precipitation occurring directly along the grain boundaries, has a more potent effect than a uniform distribution \cite{Elst1988} and it is likely that this dynamic equilibrium state will occur at a more refined grain size than indicated by the classical Zener equation.

\subsection{Precipitate growth mechanism \label{sec-precipitate-growth}}

The absence of precipitate growth, despite the continuation of dynamic precipitation is a surprising result, given the accelerated diffusion operating during \gls{hpt}.  
Precipitate growth is generally considered to operate under either diffusional or interfacial control\footnote{More strictly, bulk diffusion and diffusion across the matrix-particle interface.}.  The two mechanisms were compared by Nolfi et. al. \cite{Nolfi1970} and Radhakrishnan \cite{Radhakrishnan1993} in the context of precipitate dissolution. Although this neglects the influence of nucleation sites -- arguable abundant here given the amount of grain boundary area in a \gls{ufg} alloy -- the basic premise remains valid. This described the dissolution process by a reaction constant, $\sigma$, where:
\begin{equation}
\sigma = \left(\frac{KR_0}{D} +1 \right)^{-1}
\end{equation} and 
$K$ was an interfacial reaction constant, the $R_0$ precipitate radius and $D$ the diffusion coefficient in matrix at a given temperature.  This model predicts diffusional control for  $KR_0 \gg D$ and interface controlled for $KR_0 \ll D$. Growth of Mn particles in Mg-Mn alloy has been considered in the early work by Smith\cite{Smith1967}, who considered both diffusion and interface-controlled growth, but was not able to draw a firm conclusion.

Diffusional growth is characterised by Ostwald ripening, where the stronger diffusional field surrounding larger particles allows them to grow at the expense of smaller neighbouring particles. This results in a distinctly different particle size distribution from interfacial growth\cite{Smith1967}.

The \gls{lsw},  model,   
predicts a frequency distribution radius $r$ at a given time $t$ is given by: 
\begin{equation}
f(r,t)  = A p^2 h(p)
\end{equation}
where  $A$ depends only on the process time, which is in this case proportional to the \gls{hpt} deformation imposed, and the relative radius, $p$, is $r/\overline{r}$
and
\begin{equation}
h(p) = \left( \frac{3}{3+p} \right)^{7/3}  \left( \frac{1.5}{1.5 -p}\right)^{11/3} \exp\left( - \frac{p}{1.5-p} \right)
\end{equation}

In comparison, for interface control, the frequency shows a linear \cite{Kesternich1985} rather than quadratic dependence on $p$ \cite{Wagner1961}, with :
\begin{equation}
f(r,t) = A p h(p) 
\end{equation}
and 
\begin{equation}
h =\frac{2}{(2 -p)}^5  \exp{ \frac{-3p}{2-p }}
\end{equation}

\begin{figure}
	\centering
	\subfloat[\label{fig:rmsmodels-plot}]{\includegraphics[width=0.44\textwidth]{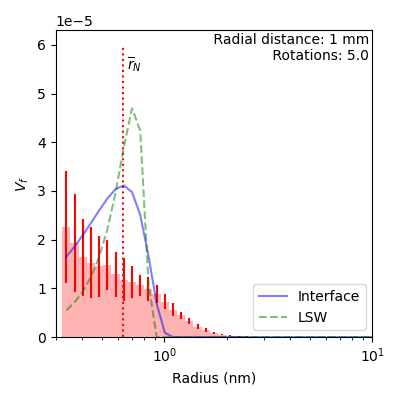}}
	
	\caption{A comparison of models of the precipitate radius distribution. The histogram shows a typical number-weighted particle radius distribution from the \gls{saxs} measurements. 
		The blue and green lines show fits to the particle size distribution for the \gls{lsw} and interfacial control models, respectively. 
	}
	\label{fig:rmsmodels}
\end{figure}

A comparison between the two models and the experimental data for 5 rotations at a radial distance of 1\,mm  is shown in Fig~\ref{fig:rmsmodels}. The number-weighted volume fraction and its uncertainty are shown by the histogram and solid vertical lines (as in Figure\ref{fig:saxsradiushistogramsb}).  The solid curve (blue in the colour version online) shows the predicted size distribution for interfacial control, calculated for the mean radius and total volume fraction in this deformation condition. Similarly, the dashed curve (green in the color version online) shows the expected particle size distribution for the \gls{lsw} model. 

While neither model provides a fully satisfactory description of the particle size distribution, the interfacial diffusion model is clearly in better agreement.  In particular, the \gls{lsw} model predicts a more monodisperse size distribution, and much lower volume fractions at radii less than the number-weighted mean than was measured. These distinguishing features were consistent for all deformation conditions, with the interfacial diffusion model better describing the experimental data in all cases.

The \gls{saxs} data does not exclude the possibility that fragmentation or shearing of Mn particles also plays a role in the retention of a fine particle size, though \gls{tem} studies found Mn particles to be resistant to shearing during \gls{hpt} of a Mn-1\%Mn-1\%Nd alloy\cite{WangCepeda2020}.

Particle fragmentation has been reported in \gls{hpt} solution-treated Al-3wt\% Cu alloys \cite{Nasedkina2017,Hohenwarter2014}. Nasedkina et. al. \cite{Nasedkina2017} argued that precipitation of elongated, aligned $\theta$  precipitates was followed by fragmentation and/or plastic deformation at higher strains. This was evidenced by significant decreases in the particle size  (from an initial value of 70 to around 40\,nm) accompanied by changes in the precipitate morphology, with a growing proportion  of relatively isotropic particles. The development of a bimodal precipitate size distribution in Al-Zn-Mg-Cu\cite{Duchaussoy2020} was also taken as an indicator of particle shearing or fragmentation.

The signatures of particle fracture or fragmentation should include not only a reduction in the particle size, and in particular the volume-weighted particle size, but also changes in the precipitate morphology. Although particle fragmentation cannot be ruled out in the Mg-Mn alloy, the minimal change in particle size and aspect ratio (Fig~\ref{fig:tem-particle-aspect-ratio}) suggests that it plays, at most, a minor role. The particle size distributions also suggest more pronounced growth of the smaller fraction, with limited changes observed for larger particle sizes. This is consistent with the small size of the particles and their resistance to shearing under conventional processing conditions. 

\subsection{Summary}	

The \gls{haadf}-\gls{stem} images suggest strongly that nucleation is restricted to the grain boundaries and triple points. \gls{hpt} deformation generally favours grain boundary precipitation, for examples in  Al-Zn-Mg-Cu\cite{Duchaussoy2020}, where this lead to a more rapid solute depletion in the matrix.
Much larger precipitates forming at grain boundaries and triple points during \gls{hpt} of Al-Zn alloys \cite{Chinh2020}, with sizes up to 150–200\,nm in Zn rich compositions (10-30wt.\%). However, in both cases intragranular precipitation was still observed.

The particle size distributions are more consistent with interfacially rather than diffusionally controlled growth. This explains the lack of substantial particle growth despite  the accelerated diffusion during \gls{hpt}  and the close proximity of many particles. Whether this is a general feature of precipitation of Mn in Mg is not clear, but it has long been noted that a diffusional model for growth of Mn particles in isothermal condition would have to impose an anomalously low frequency factor to fit the experimental  observations \cite{Smith1967}.

Given the ultrafine grain size and absence of obvious grain boundary segregation, the resistance to grain growth is attributed to particle pinning. \gls{saxs} data indicates that this precipitation continues to at least 10 rotations of deformation, and that there remains a sizeable amount of Mn in solution to continue this process. Despite their fine size and initial low volume fraction, the effectiveness of these particles would be enhanced by their localisation at the grain boundaries and their apparent resistance to coarsening and fragmentation.

This works also highlights how a combination of \gls{tem} and \gls{saxs} can provide a more complete description of the precipitate process than either technique could alone\cite{RosaliePauw2014, Duchaussoy2020}. Atomic-contrast \gls{stem} and \gls{eds} provided the spatial distribution, morphology and chemistry of the particles. Obtaining accurate particle-sizes measurements from statistically-relevant sample populations is problematic\cite{RosaliePauw2014, Duchaussoy2020}. \gls{saxs} was far-better suited to providing accurate particle-size distributions and particle-volume fractions, but required the morphological information from \gls{tem} to inform the model. 

Extended deformation of the \gls{ufg} Mg-Mn appears unlikely to alter the behaviour already apparent after 10 rotations, or an equivalent strain $>$300\cite{Rosalie2025}. The relative stability of the precipitate size distribution, and the continued existence of a large solute reservoir suggest that additional precipitation, coupled with grain growth, can continue until a point is reached where the particle density is sufficient to completely stabilise the grain size.

However, the presence of residual solute suggests that precipitation could also be encouraged via suitable heat treatment. Equation~\ref{fig:zenerpinning} indicates that an increase in the particle radius would also promote more effective pinning of the grain boundaries. Whether this enhanced pinning would occur rapidly enough to counteract grain growth at the same temperature remains to be answered. 

\section{Conclusions}

Nanometer-scale particles precipitated on grain boundaries during room temperature \gls{hpt} of solutionised Mg-1.35wt.\%.Mn. Compositional mapping indicated that Mn was localised in the particles only, and  generalised grain-boundary segregation was not observed. This was attributed to the ultrafine grain size, which meant that there was insufficient Mn to populate the grain boundary surfaces. 

Dynamic precipitation proceeded sluggishly during \gls{hpt}, with only $\sim$24\% of the Mn partitioned to nanometer-scale grain boundary particles after 10 rotations. This appears to derive from a relatively low diffusivity for Mn in Mg, even with the accelerated diffusion operating during \gls{spd}.

While the extent of precipitation was limited by diffusion, the growth of individual particles was not. The particle size distribution, and the continued existence of grain boundary particles in close proximity, suggest that the growth of the particles is subject to interfacial control. This would explain the relatively constant mean particle size over extended periods of deformation. While it is not possible to rule particle fragmentation, there was no direct evidence to support its occurrence and the Mn particles are known from previous reports to be resistance to shearing. 
 
The magnesium grains are stabilised via particle pinning, allowing finer grain sizes to be acheived than in pure Mg. The amount and size of particles is sufficient to retard, but not completely halt, grain growth, explaining the gradual increase in grain size during \gls{hpt}. Further deformation is expected to result in a) additional precipitation of 1-2\,nm radius Mn particles, and b) grain coarsening, until a dynamic equilibrium between these effects is reached. 

\section*{Data availability}
Data will be made available on reasonable request.



\section*{Acknowledgements}
The authors would like to thank Dr. Martyn Alderman of Magnesium Elektron for providing the material used in this investigation. Ion polishing of the \gls{tem} samples was assisted by C. F\"orster, (Helmholtzentrum, Berlin). The \gls{sem} observations were conducted by P. Ocano-Suarez (\gls{bam}). \gls{haadf}-\gls{stem} measurements were carried out with the assistance of C. Prinz (\gls{bam}). 

\section*{Author contributions}

J.M.R. Designed the study, performed the TEM experiments, analysed the corresponding data and wrote the main manuscript.
A.H. Prepared the \gls{hpt} samples, performed the heat treatments and conducted the deformation experiments. 
B.R.P performed the \gls{saxs} experiments and analysed the corresponding data.
All authors reviewed the manuscript.

\section*{Ethics declarations}
\subsection*{Ethics, consent to participate, and consent to publish}
Not applicable.

\subsection*{Competing interests}
The authors declare no competing interests.

\end{document}